\documentclass[aps,preprint,superscriptaddress,pra]{revtex4-1}

\usepackage{amsmath,amssymb,graphicx,color,verbatim,soul,braket}

\usepackage{hyperref}
\usepackage[normalem]{ulem}

\begin{document}

\title{Experimental extraction of the quantum effective action for a non-equilibrium many-body system}

\author{Maximilian Pr\"ufer}
\thanks{These authors contributed equally to this work}
\affiliation{Kirchhoff-Institut f\"ur Physik, Universit\"at Heidelberg, Im Neuenheimer Feld 227, 69120 Heidelberg, Germany}
\author{Torsten V. Zache}
\thanks{These authors contributed equally to this work}
\affiliation{Institut f\"ur Theoretische Physik, Universit\"at Heidelberg, Philosophenweg 16, 69120 Heidelberg, Germany}
\author{Philipp Kunkel}
\affiliation{Kirchhoff-Institut f\"ur Physik, Universit\"at Heidelberg, Im Neuenheimer Feld 227, 69120 Heidelberg, Germany}
\author{Stefan Lannig}
\affiliation{Kirchhoff-Institut f\"ur Physik, Universit\"at Heidelberg, Im Neuenheimer Feld 227, 69120 Heidelberg, Germany}
\author{Alexis Bonnin}
\affiliation{Kirchhoff-Institut f\"ur Physik, Universit\"at Heidelberg, Im Neuenheimer Feld 227, 69120 Heidelberg, Germany}
\author{Helmut Strobel}
\affiliation{Kirchhoff-Institut f\"ur Physik, Universit\"at Heidelberg, Im Neuenheimer Feld 227, 69120 Heidelberg, Germany}
\author{J\"urgen Berges}
\affiliation{Institut f\"ur Theoretische Physik, Universit\"at Heidelberg, Philosophenweg 16, 69120 Heidelberg, Germany}
\author{Markus K. Oberthaler}
\affiliation{Kirchhoff-Institut f\"ur Physik, Universit\"at Heidelberg, Im Neuenheimer Feld 227, 69120 Heidelberg, Germany}

\date{\today}

\maketitle
\textbf{
Far-from-equilibrium situations are ubiquitous in nature. They are responsible for a wealth of phenomena, which are not simple extensions of near-equilibrium properties, ranging from fluid flows turning turbulent to the highly organized forms of life \cite{Equ2007}. 
On the fundamental level, quantum fluctuations or entanglement lead to novel forms of complex dynamical behaviour in many-body systems \citep{polkovnikov2011} for which a description as emergent  phenomena can be found within the framework of quantum field theory.
A central quantity in these efforts, containing all information about the measurable physical properties, is the quantum effective action \cite{Weinberg1995}.
Though the problem of non-equilibrium quantum dynamics can be exactly formulated in terms of the quantum effective action, the solution is in general beyond capabilities of classical computers \citep{Jordan2011}.
In this work, we present a strategy to determine the non-equilibrium quantum effective action \citep{Wetterich1997} using analog quantum simulators, and demonstrate our method experimentally with a quasi one-dimensional spinor Bose gas out of equilibrium \citep{Sadler2006a,StamperKurn2013}.
Building on spatially resolved snapshots of the spin degree of freedom \citep{Kunkel2019}, we infer the quantum effective action up to fourth order in an expansion in one-particle irreducible correlation functions at equal times. 
We uncover a strong suppression of the irreducible four-vertex emerging at low momenta, which solves the problem of dynamics in the highly occupied regime far from equilibrium where perturbative descriptions fail \citep{pitaevskii2012}.
Similar behaviour in this non-pertubative regime has been proposed in the context of early-universe cosmology \citep{Berges2008}.
Our work constitutes a new realm of large-scale analog quantum computing \citep{Horsman2014}, where the high level of control of synthetic quantum systems \citep{Bloch2012} provides the means for the solution of long-standing theoretical problems in high-energy and condensed matter physics with an experimental approach \citep{zohar2015,Lewenstein2007,loh1990}.
}

In the many-body limit the measurable physical properties of an interacting quantum system are determined by only a small subset of all microscopic parameters.
As a consequence of this effective loss of details, very efficient descriptions of quantum many-body systems can be found using a concept known as the renormalization program of quantum field theory \citep{Weinberg1995,Wilson1975}.
Renormalization implies a scale-dependent description that links the physics on small characteristic distances with phenomena on macroscopic length scales. 
The scale dependence is encoded in `running' couplings, defined as momentum dependent expansion coefficients of the quantum effective action. 
Matching the experimental capabilities, we choose a formulation of quantum field theory based on equal-time correlation functions only \citep{Wetterich1997}.
With this, finding the time-dependent quantum effective action $\Gamma_t$, equivalent to solving the non-equilibrium dynamics, involves experimentally observable expectation values of the underlying quantum fields.
Therefore, the problem is mapped onto the ability of synthetic quantum systems giving higher-order correlation functions~\citep{Schweigler2017,Hodgman2017,Rispoli2018,Preiss2019}.

Our quantum simulation builds on a spinor Bose-Einstein condensate \citep{StamperKurn2013} and the spatially resolved detection of a complex-valued field $F_\perp(y) = F_x(y) + \text{i}F_y(y) = |F_\perp(y)|\text{e}^{i\varphi(y)}$ \citep{Kunkel2019} (see Methods).
We employ $\sim 100,000$ $^{87}$Rb atoms in a quasi one-dimensional regime and quench an experimental control parameter that leads to spin dynamics in the $F = 1$ hyperfine manifold (see Fig.~1a), building up excitations in the $F_x$-$F_y$-plane \citep{Kawagushi2012}. 
Here, $F_x$ and $F_y$ are the components of the hyperfine spin perpendicular to an applied external magnetic field.
With an optimized optical trap we achieve coherence times up to $\sim50\,$s due to reduced heating and efficient evaporative cooling.
We find that the dynamics leads to an approximately constant spin length $|F_\perp|$ and a fluctuating phase degree of freedom $\varphi$ (see Fig.~1c), resembling the structure of a single-component Bose gas.
By averaging over many realizations, we infer an estimator of the two-~and four-point correlation functions (correlators) from the single-shot results of $F_\perp(y)$ (see Fig.~1b and Methods \textbf{B}).

\begin{figure}
    \linespread{1}
	\centering
	\includegraphics[width = 0.75\columnwidth]{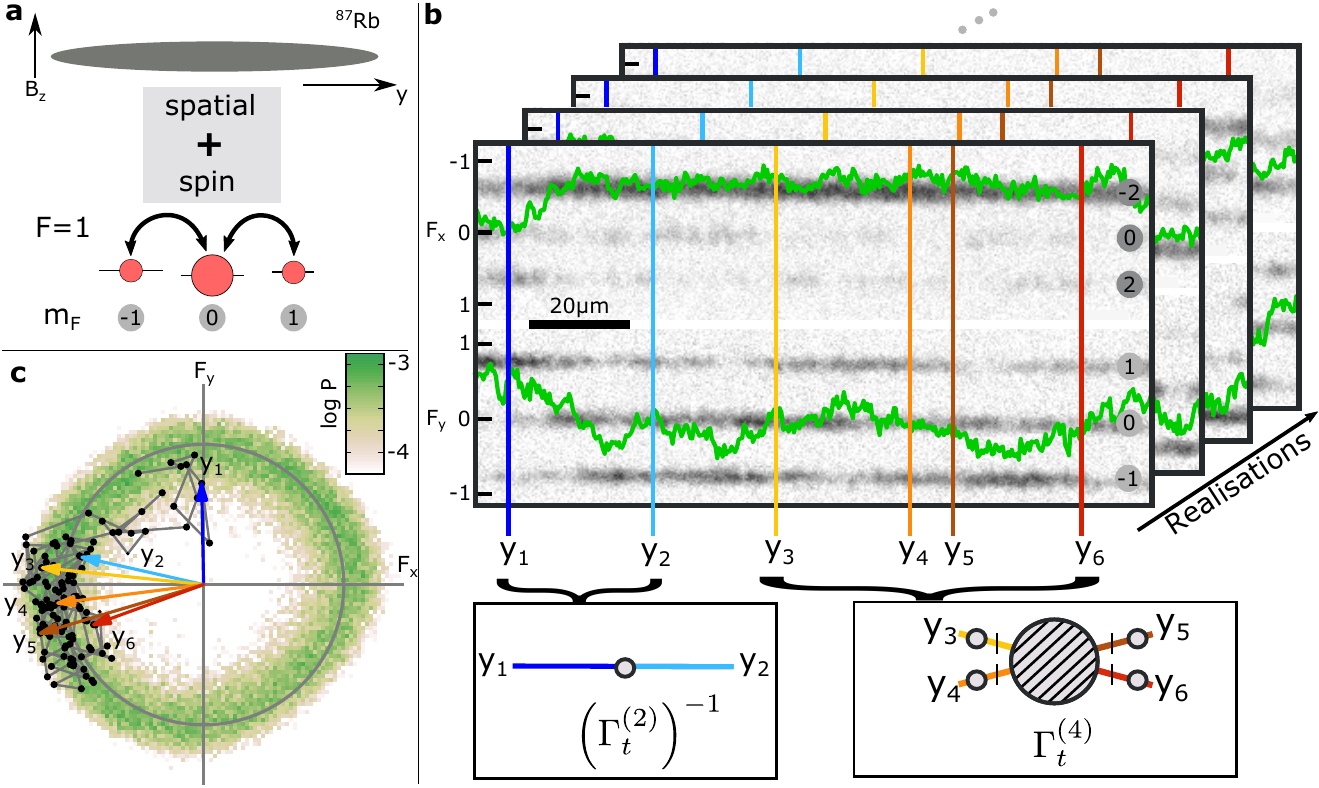}
	\caption{\textbf{Experimental platform and extraction of correlation functions.} \textbf{a,} The experimental system consists of an elongated, quasi one-dimensional $^{87}$Rb Bose-Einstein condensate in a magnetic field $B_z$ in z-direction. The spin dynamics takes place in the $F = 1$ hyperfine manifold. \textbf{b,} In a single realization we access spatially resolved snapshots of the dynamics. We infer the transversal spin  $F_y$ and $F_x$ (green lines) from atomic densities (grey shading) measured after a spin mapping sequence in the indicated magnetic substates in the F = 1 and F = 2 hyperfine manifolds, respectively (for details see methods). Many realizations are used to determine the correlation functions of $F_\perp = F_x + iF_y$ to infer the proper vertices $\Gamma^{(n)}_t$. \textbf{c,} The distribution of $F_\perp$ (all realizations and all spatial points, see Ext.\,Data Fig.\,1 for all evolution times) for 18\,s evolution time. The black dots indicate the single realization shown in \textbf{b}; neighbouring points are connected by a grey line. The coloured arrows correspond to the spin inferred at the points $y_1,...,y_6$.}
	\label{Distribution}
\end{figure}

At evolution time $t$, the corresponding quantum effective action $\Gamma_t$ (see Methods for the connection to the Wigner distribution, e.g., used in quantum optics \citep{Hillrey1984}) is a functional depending on the macroscopic field $\Phi^1(y)$ and its canonically conjugated field $\Phi^2(y)= \left( \Phi^1(y)\right)^*$.
Using a compact notation, the full quantum effective action may then be written as an expansion in the fields as (see Methods)
\begin{align}\label{eqn:qea}
\Gamma_t[\Phi] &=  \sum_{n=1}^{\infty} \frac{1}{n!}    \, \Gamma_t^{\alpha_1, \dots, \alpha_n}(y_1, \ldots, y_n) \, \Phi^{\alpha_1}(y_1) \cdots \Phi^{\alpha_n}(y_n)   , 
\end{align}
where we sum over repeated indices, $\alpha_j=1,2$ and integrate over all coordinates $y_j$ with $j=1,\dots,n$. 
The expansion coefficients $\Gamma_t^{\alpha_1, \dots, \alpha_n}(y_1, \ldots, y_n) $ in  (1), so-called proper vertices \citep{Weinberg1995}, are the one-particle irreducible (1PI) $n$-point correlation functions of $F_\perp(y)$. 
By assuming $U(1)$ symmetry for our system, the only independent coefficients are $ \Gamma_t^{1\dots12\dots2}(y_1,\dots,y_n) \equiv \Gamma_t^{(n)}(y_1,\dots,y_n)$ with $n$ even and equally many field and conjugate field components.
These quantities directly characterize the propagation ($\,\Gamma_t^{(2)}\,$) and interactions ($\,\Gamma_t^{(n>2)}\,$) of the field by taking into account the effects of all quantum-statistical fluctuations.
Any physical observable may be constructed from the knowledge of all proper vertices at all times since they contain the same information as the density matrix (see Methods \textbf{E}).

\begin{figure}
    \linespread{1}
	\centering
	\includegraphics[width = 0.5\columnwidth]{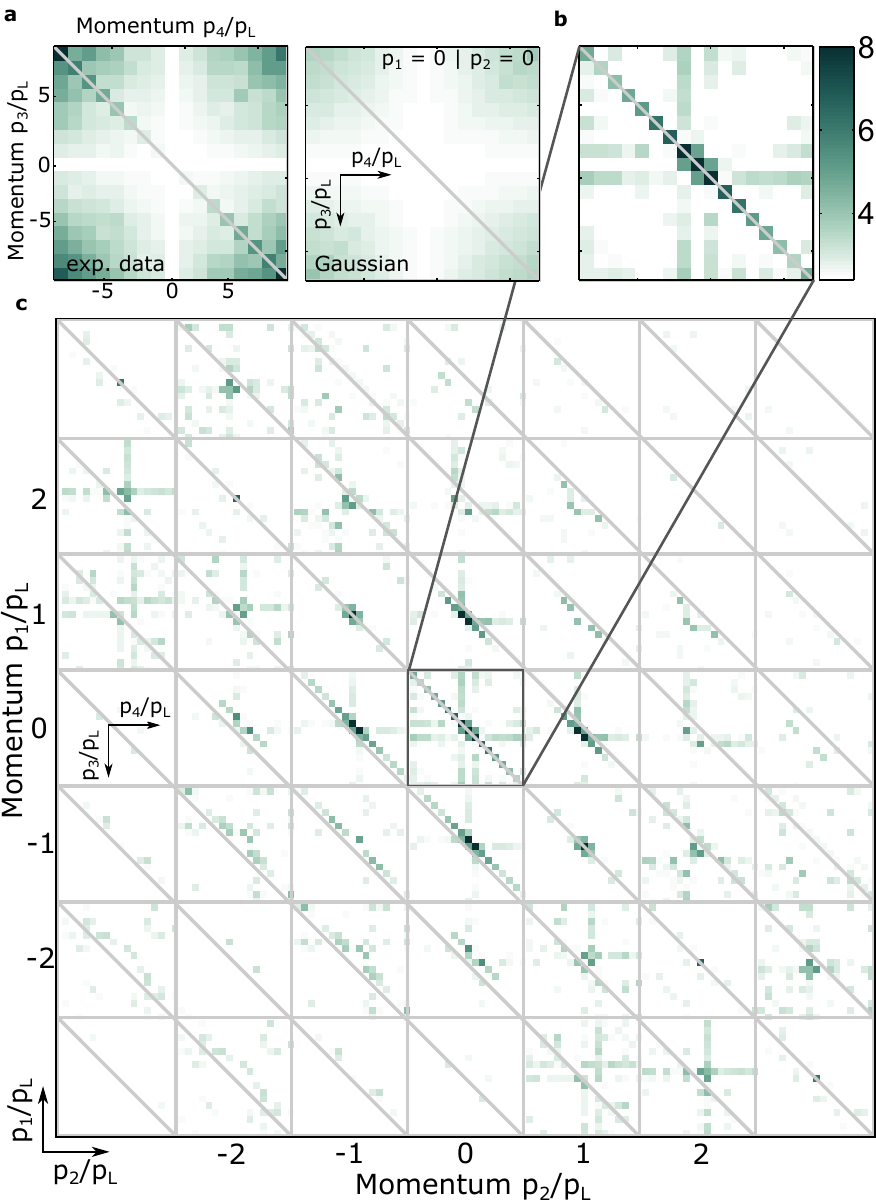} 
	\caption{\textbf{Statistical significance of the four-point 1PI correlator in momentum space.} \textbf{a,} 1PI four-vertex $\Gamma^{(4)}_{t=18s}(0,0,p_3,p_4)$ for experimental data (left panel); for Gaussian model with same statistics (right panel). $p_L = 1/L$ is the lowest momentum corresponding to the size $L$ of the evaluation region. \textbf{b,} Ratio of experimental data and Gaussian model indicating the statistical significance of the inferred signal. \textbf{c,} $\Gamma^{(4)}_{t=18s}(p_1,p_2,p_3,p_4)$ for different $p_1$ and $p_2$ (fixed in each grey square), divided by the Gaussian bias, revealing an overall momentum conserving structure.}
	\label{Distribution}
\end{figure}

The proper vertices represent the irreducible building blocks for the description of the quantum many-body dynamics, and they can be related to the full correlation functions measured such as $\braket{F_\perp(y_1)F_\perp^*(y_2)}_{t}$ or $\braket{F^*_\perp(y_1)F_\perp^*(y_2)F_\perp(y_3)F_\perp(y_4)}_{t}$. 
We subtract the redundant disconnected parts from the full correlators to obtain the connected correlation functions $C_t^{(2)}(y_1,y_2) = \braket{F_\perp(y_1)F_\perp^*(y_2)}_{t,\text{c}}$ and $C^{(4)}_t(y_1,y_2,y_3,y_4) =  \braket{F^*_\perp(y_1)F_\perp^*(y_2)F_\perp(y_3)F_\perp(y_4)}_{t,\text{c}} $ \citep{Schweigler2017}. 
The connected correlators are then decomposed into their irreducible parts representing the proper vertices (see Methods).
We obtain a momentum resolved picture by performing a discrete Fourier transform, which yields $|C_t^{(n)}(p_1, \ldots, p_n)|$ with momenta $p_j$ (see Methods). 
With that the macroscopic, long wavelength, behaviour of the quantum system is encoded in the low-momentum or ``infrared'' properties of the 1PI correlation functions.
More precisely, we extract $\Gamma^{(2)}_t(p,-p) \equiv \Gamma^{(2)}_t(p) = \left(C_t^{(2)}\right)^{-1}(p)$ from the inverse of the connected two-point correlator and with that the 1PI four-point correlation function for our $U(1)$-symmetric case: 
\begin{equation}
\Gamma^{(4)}_t(p_1,p_2,p_3,p_4) = -\Gamma^{(2)}_t(p_1)\,\Gamma^{(2)}_t(p_2)\,\Gamma^{(2)}_t(p_3)\,\Gamma^{(2)}_t(p_4)\,C^{(4)}(p_1,p_2,p_3,p_4)\, .
\end{equation}
Pictorially, acting with the inverse two-point correlators on the connected four-point function removes the `external legs' of the diagram (see Fig.~1b). 

The experimental extraction of the 1PI four-point correlations requires sufficiently many experimental realizations to ensure the statistical significance of the results.
In Fig.~2a we plot a slice of $\Gamma^{(4)}_{t=18s}(p_1,p_2,p_3,p_4)$ by fixing $p_1 = p_2 = 0$ and showing its dependence on $p_3$ and $p_4$ (see Ext.~Data Fig.~2).
The correlator exhibits an overall momentum dependence with a prominent contribution on the momentum conserving diagonal ($p_1+p_2+p_3+p_4=0$).
The values found are contrasted to a finite statistical bias obtained from a Gaussian model with the same number of realizations as in the experiment (see Methods).
The momentum-conserving diagonal is shown to be statistically significant by dividing the values inferred from the experimental data by the finite statistical bias (see Fig.~2b).
In Fig.~2c the same proper vertex divided by the Gaussian bias is shown  for different $p_1$ and $p_2$.
This corroborates the overall momentum conserving structure. 

\begin{figure}
    \linespread{1}
	\centering
	\includegraphics[width = 1\columnwidth]{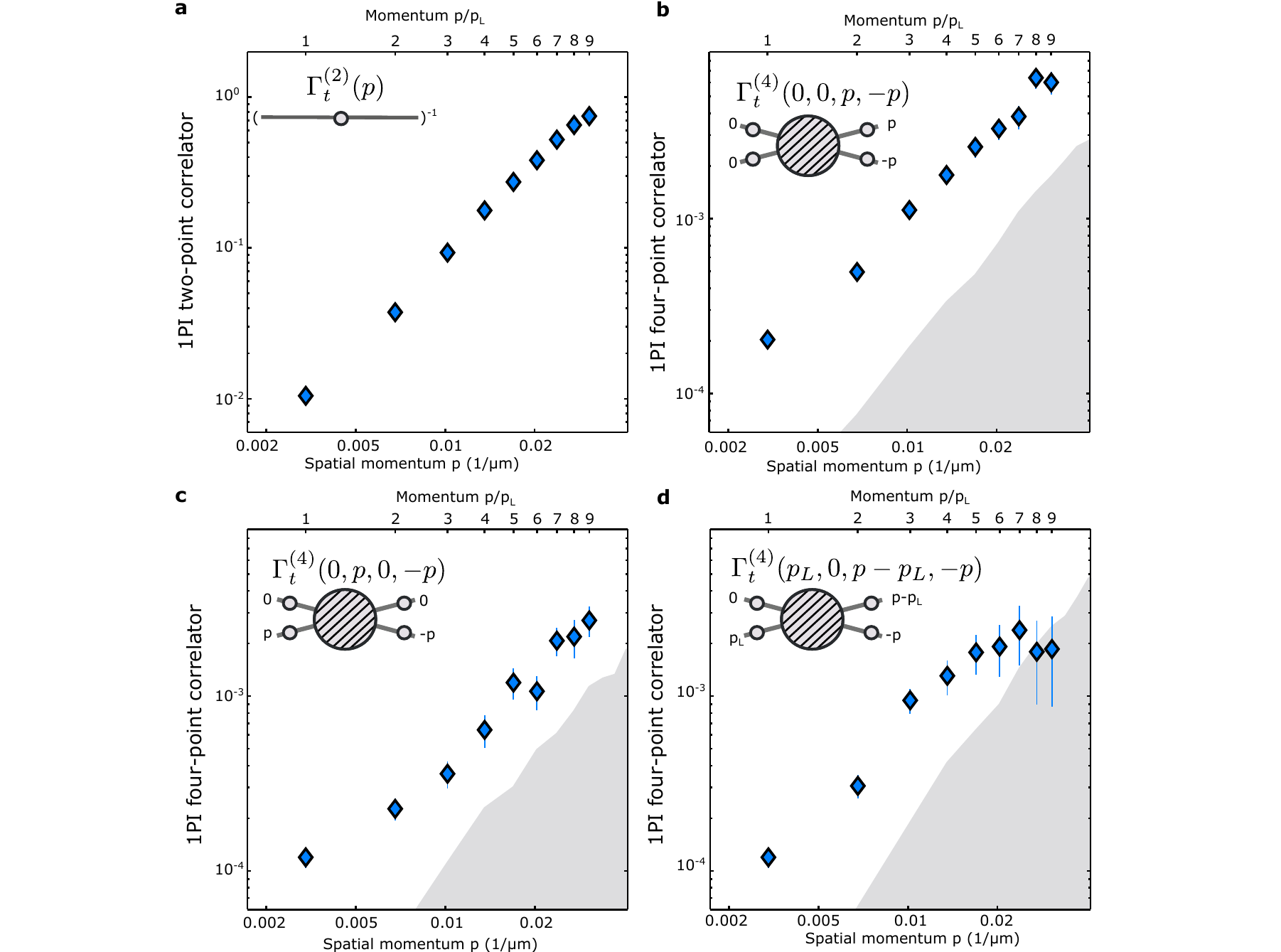} 
	\caption{\textbf{Momentum conserving diagonals of the 1PI correlators.} \textbf{a,} Momentum conserving diagonal of $\Gamma^{(2)}_{t=18s}$. \textbf{b-d,} We show three different momentum conserving diagonals of the four point 1PI correlations (blue diamonds) for $t = 18\,$s (see Ext. Data Figure 3 for all evolution times). They are statistically significant as quantified by the comparison to the finite statistics level for the Gaussian case (grey shaded region, see Methods). The complex-valued field allows the distinction between case \textbf{b} and \textbf{c}. The momentum dependence of these correlators over almost two orders of magnitude indicates the dramatic renormalisation of the couplings.  All error bars shown are 1~s.d. calculated from bootstrap resampling.}
	\label{Distribution}
\end{figure}

For quantifying the renormalization effects, we investigate the momentum dependence of the proper vertices.
Fig.~3a  shows $\Gamma^{(2)}_{t=18s}(p)$ as a function of the spatial momentum $p$. 
We observe a suppression in the infrared \citep{Pruefer2018,Erne2018} up to a characteristic momentum set by the inverse of the length scale $l_s\sim 30\,\mu$m.
For $\Gamma^{(4)}_{t=18s}(p_1,p_2,p_3,p_4)$, we focus on statistically significant momentum conserving diagonals.
As a representative example, the case of  $p_1 = p_2 = 0$ and $p = p_3 = -p_4$ (Fig.~3b) shows strongly momentum-dependent values in the observed momentum regime; a similar behaviour is also found for different diagonals as shown in Fig.~3c and 3d.
This demonstrates that the infrared regime exhibits a strongly suppressed interaction vertex in this far-from-equilibrium situation \citep{Berges2008,Orioli2015,Walz2017,Chantesana2018a}.

Motivated by the observed structure of the complex valued field, we derive an exact evolution equation for the two-point function $C_t^{(2)}(p)$ of a one-component Bose gas with interaction strength $g$ (assuming spatial translational invariance; see Methods):
\begin{align}\label{eq:fullBoltzmann}
\partial_t C_t^{(2)}(p) &= g \int_{q,r,l}  i\left[\Gamma_t^{(4)}(p,q,-r,-l) - \Gamma_t^{(4)}(l,r,-q,-p) \right]\, C_t^{(2)}(p) C_t^{(2)}(q) C_t^{(2)}(r) C_t^{(2)}(l) \;.
\end{align}
Here, $C_t^{(2)}(p)$ has the interpretation of an occupation number distribution, where the total number $\int_p C_t^{(2)}(p)$ is conserved because of the $U(1)-$symmetry assumed. 
Eq.~(3) may be viewed as the full quantum field theoretical version of kinetic descriptions for the time evolution of $C_t^{(2)}(p)$.
With our results we have determined the defining parameters $\Gamma_t^{(4)}(p_1,p_2,p_3,p_4)$ of this exact evolution equation, answering the long-standing question about the dynamics in the highly occupied regime, which cannot be captured by standard kinetic theory.
Fig.~3 and 4 show that the momentum-dependent vertex $\Gamma_t^{(4)}$ drops by almost two orders of magnitude as the occupancy grows strongly towards lower momenta. 
The reduced effective interaction strength diminishes the rate of change of the distribution $C^{(2)}(p)$ according to (\ref{eq:fullBoltzmann}), counter-acting the Bose-enhancement from the high occupancies at low momenta.  

Moreover, in our experiment we find self-similar dynamics for the statistically accessible diagonals of $\Gamma_t^{(4)}$ according to $\Gamma^{(4)}_{t}(0,0,p,-p) =  t^\gamma\Gamma_S(0,0,t^{\beta_4} p,-t^{\beta_4} p)$ with scaling exponents $\gamma$ and $\beta_4$.
In Fig.~4b we show the time evolution (inset) and the rescaled data with $\gamma = 0$ and $\beta_4= 1/2$ revealing a scaling collapse on the function $\Gamma_S \propto 1+(p/p_S)^{\zeta_4}$ with $\zeta_4=2.2$.
We also find scaling of the two-point correlator $C_t^{(2)}= t^{\alpha} C_{S}(t^{\beta_2} p)$ with $C_S \propto \left( 1+(p/p_S)^{\zeta_2} \right) ^{-1}$ and the expected exponents $\alpha = \beta_2 \simeq 1/2$ \citep{Orioli2015,Pruefer2018} (see Fig.~4a).
The evolution equation (3) connects $(\alpha,\beta_2)$ to $(\gamma,\beta_4)$.
Testing a relation of $ \alpha$ and $\gamma$, implicitly given by the integration in Eq. (3), is currently hampered by the limited statistics.
We anticipate that $C_t^{(2)}$ and $\Gamma^{(4)}_{t}$ have the same momentum scaling exponents, i.e.~$\beta_2 = \beta_4$, which is consistent with our findings.

\begin{figure}
    \linespread{1}
	\centering
	\includegraphics[width = 1\columnwidth]{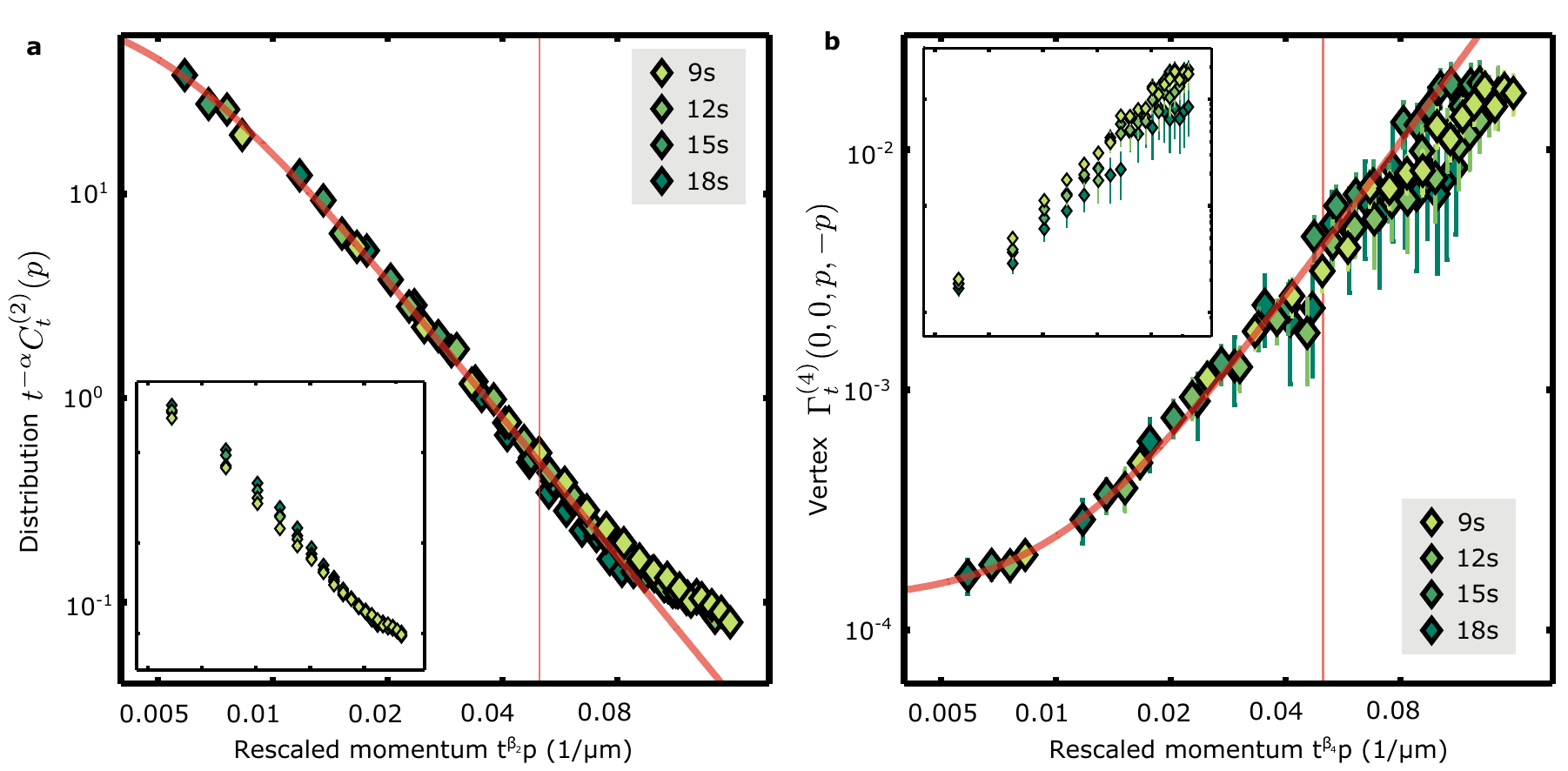} 
	\caption{\textbf{Observation of scaling in time of the distribution function as well as corresponding couplings.} Rescaled distribution function $C_t^{(2)}(p)$ and 1PI four point correlators $\Gamma_t^{(4)}(0,0,p,-p)$ for times between 9\,s and 18\,s evolution time (see Ext.~Data Fig.~4 for the other diagonals shown in Fig.~3). We fit a universal function (red line; see text) with power law $\zeta$ and scale $p_S$ up to the indicated momentum scale (red vertical line). The inset shows the unscaled data with same axis ticks as the main figure. \textbf{a}, We use $\alpha = \beta_2=1/2$ for rescaling and find $p_S = 1/199\, \mu \text{m}^{-1}$ and $\zeta_2 = 2.3$. Here, error bars are smaller than the size of the plot markers. \textbf{b}, We use $\beta_4=1/2$ for rescaling and find $p_S = 1/93\, \mu \text{m}^{-1}$ and $\zeta_4 = 2.2$. All error bars shown are 1 s.d. calculated from bootstrap resampling.}
\end{figure}

While our analog quantum simulator employs a specific setup with a spinor Bose gas, many of the results are insensitive to the detailed properties of the device on short-distance scales because of renormalization. 
The latter makes analog quantum simulators in the many-body limit versatile, with possible quantitative applications to a wide range of systems described by quantum field theory.

\newpage

\vspace*{0.5cm}
\begin{center}
  \textbf{Methods}
\end{center}
\vspace*{0.5cm}

\subsection{Experimental details}
In the experiment, we prepare $\sim10^5$ atoms in the state $\ket{\text{F},m_\text{F}} = \ket{1,0}$ in a static magnetic field of $0.884\,$G in an optical dipole trap of $1030\,$nm light with trapping frequencies $(\omega_\parallel,\omega_\perp) \approx 2\pi\times(1.6,167)\,$Hz.
We initiate dynamics by changing the experimental control parameter $q$, the second order Zeeman shift, by applying off-resonant microwave dressing. 
For details on the experiment and the parameter regime employed see \citep{Pruefer2018}.

We use a readout scheme to detect multiple spin projections in a single realization \citep{Kunkel2019}.
After the desired evolution time $t$ we apply a first $\pi/2$-rf rotation around the y-axis to map the spin projection $F_x$ on the detectable population imbalance $N_{+1}-N_{-1}$. 
We store this projection in the initially empty $F = 2$ hyperfine manifold by splitting the populations of the three $m_F$ states with three mw $\pi/2$ pulses. 
A second, hyperfine selective, rf rotation in $F = 1$ around the $x$-axis allows us to map the spin projection $F_y$ onto the population imbalance in $F = 1$.
Together with a single absorption image of the 8 hyperfine levels this procedure allows us to extract single shot snapshots of the complex valued order parameter $F_\perp(y) = F_x(y)+iF_y(y)$ with 
\begin{align}
F_x(y) &= \left( N_{+2}^{F=2}(y) - N_{-2}^{F=2}(y) \right) / N_\text{tot}^{F=2}(y)\\
F_y(y) &= \left( N_{+1}^{F=1}(y) - N_{-1}^{F=1}(y) \right) / N_\text{tot}^{F=1}(y),
\end{align}
where $N_{m_f}^{F=f}(y)$ is is the atom number in the state $\ket{f,m_f}$ at position $y$ and $N_\text{tot}^{F=f}(y)$ is the total atom number in the hyperfine manifold $f$ after the readout sequence.

At evolution times of $18\,$s we observe no substantial loss of coherence of the different $m_\text{F}$ states which would correspond to reductions of the transversal spin length. 
Optimizing the transversal confinement leads to coherence times up to $\sim50\,$s due to reduced heating and efficient evaporative cooling.
Due to the harmonic confinement the atomic density is inhomogeneous and so is the transversal spin length profile. 
We choose to analyse the central $400\,$pixels corresponding to $\sim168\,\mu$m. 
The spatial resolution is $\sim1.1\,\mu$m corresponding to three pixels (pixelsize corresponds to $420\,$nm). 
As we are interested in the long wavelength (infrared) excitations we sum over $9$ adjacent pixels to reduce the number of points and the computation time.
We use the number of realizations given in Table 1 for estimating the correlators.

\subsection{Explicit expressions for correlators}
Experimentally, we obtain $N$ measurements of an observable $\mathcal{O}_a$, i.e. data sets\\ $\left\lbrace \mathcal{O}^{(i)}_a \big| i = 1, \dots, N \right\rbrace$ from which we infer the $n$-th order correlation function via
\begin{align}
	\langle \mathcal{O}_{a_1} \cdots \mathcal{O}_{a_n} \rangle \approx \frac{1}{N} \sum_{i=1}^{N} \mathcal{O}^{(i)}_{a_1} \cdots \mathcal{O}^{(i)}_{a_n} \; .
\end{align}
In our case $a$ is a collective index for space and field components, e.g. for $a=(\alpha,x)$ and $\mathcal{O} = \Phi$, then $\mathcal{O}_a = \Phi^\alpha (x)$.

Formally, the information contained in all correlation functions, $\langle \mathcal{O}_{a_1} \cdots \mathcal{O}_{a_n} \rangle$ for all $n$, is equivalently stored in $Z[J]$, which is the generating functional for these correlators:
\begin{align}
	\left.\frac{\delta^n Z[J]}{\delta J_{a_1} \cdots \delta J_{a_n}}\right|_{J=0} = \langle \mathcal{O}_{a_1} \cdots \mathcal{O}_{a_n} \rangle \; .
\end{align}
Here $Z$ depends on so-called sources $J_{a}$ corresponding to the observable $\mathcal{O}_a$.
The relevant definition of $Z$ for the non-equilibrium setting is given in Methods \textbf{E}.

Connected and 1PI correlation functions are obtained by introducing additional functionals $E[J] = \log Z[J]$ and $
	\Gamma[O] = -\log Z[J(O)] + J_a(O) O_a$, 
which is the functional Legendre transformation of $E[J]$.
One finds $J(O)$ by inverting the equation $O(J) = \frac{\delta E}{\delta J}$ and $O_a$ is given in equation (9a).
Here and in the following, we use Einstein summation convention for the collective index $a$.
Explicitly, up to fourth order, the connected correlators are given by
\begin{subequations}
\begin{align}
O_a &= \left.\frac{\delta E}{\delta J_a}\right|_{J=0} = \langle \mathcal{O}_a \rangle\,, \\
C_{ab} &= \left.\frac{\delta^2 E}{\delta J_a \delta J_b}\right|_{J=0} =  \langle \mathcal{O}_a \mathcal{O}_b \rangle - O_a O_b\,,\\
C_{abc} &= \left.\frac{\delta^3 E}{\delta J_a \delta J_b \delta J_c}\right|_{J=0} =  \langle \mathcal{O}_a \mathcal{O}_b  \mathcal{O}_c \rangle -  C_{ab} O_c -  C_{bc} O_a -  C_{ca} O_b + 2 O_a O_b O_c  \,,\\
C_{abcd} &= \left.\frac{\delta^4 E}{\delta J_a \delta J_b \delta J_c \delta J_d}\right|_{J=0} = \langle \mathcal{O}_a \mathcal{O}_b  \mathcal{O}_c \mathcal{O}_d \rangle  -  C_{abc} O_d -  C_{cbd} O_a -  C_{cda} O_b -  C_{dab} O_c  \nonumber\\
&\qquad\qquad\qquad\qquad\qquad\qquad -  C_{ab} C_{cd} -  C_{ac} C_{db} -  C_{ad} C_{bc} - 6 O_a O_b O_c O_d \nonumber\\
&+ 2 C_{ab} O_{c} O_{d} + 2 C_{ac} O_{d} O_{b} + 2 C_{ad} O_{b} O_{c} + 2 C_{bc} O_{d} O_{a} + 2 C_{bd} O_{a} O_{c} + 2 C_{cd} O_{a} O_{b} \; .
\end{align}
\end{subequations}
The 1PI correlators are given by
\begin{subequations}
	\begin{align}
	\Gamma_a &= \left.\frac{\delta \Gamma}{\delta O_a}\right|_{J=0} = 0 \,, \\
	\Gamma_{ab} &= \left.\frac{\delta^2 \Gamma}{\delta O_a \delta O_b}\right|_{J=0} = \left[\left(\left.\frac{\delta^2 E}{\delta J \delta J}\right|_{J=0}\right)^{-1}\right]_{ab}\,,\\
	\Gamma_{abc} &= \left.\frac{\delta^3 \Gamma}{\delta O_a \delta O_b \delta O_c}\right|_{J=0} = -\Gamma_{aa'} \Gamma_{bb'} \Gamma_{cc'}C_{a'b'c'}\,, \\
	\Gamma_{abcd} &= \left.\frac{\delta^4 \Gamma}{\delta O_a \delta O_b \delta O_c \delta O_d}\right|_{J=0} =  - \Gamma_{aa'} \Gamma_{bb'} \Gamma_{cc'} \Gamma_{dd'} C_{a'b'c'd'} \nonumber\,,\\
	&\qquad\qquad + \Gamma_{aa'} \Gamma_{bb'} \Gamma_{cc'} \Gamma_{dd'} \left( C_{a'b'e} \Gamma_{ef} C_{fc'd'} + C_{a'c'e} \Gamma_{ef} C_{fb'd'} + C_{a'd'e} \Gamma_{ef} C_{fc'b'}\right)
	\end{align}
\end{subequations}
They are the expansion coefficients of the full quantum effective action in a functional Taylor expansion,
\begin{align}
\Gamma[O] = \Gamma_{ab}O_aO_b +\Gamma_{abc}O_aO_bO_c+\Gamma_{abcd}O_aO_bO_cO_d+\dots
\end{align}
In practice, we calculate the connected correlation functions in position space using the julia package \textit{Cumulants.jl}~\cite{Domino2017}. 

We expect the system to be $U(1)$ symmetric, which renders all correlators involving unequal number of $F_\perp$ and $F_\perp^*$ zero in the infinite statistics limit. For the calculation of the 1PI four-vertex, we neglect the connected three-point functions accordingly.
Additionally, we checked that the two-point functions are approximately translation invariant, which further reduces computation time by calculating the 1PI correlators directly in Fourier space.

\subsection{Momentum resolved picture}
We calculate all connected correlators in position space and find translation invariance.
To obtain a momentum resolved picture we perform a discrete Fourier transform:
\begin{align}
C_t^{\alpha_1,\dots,\alpha_n}(p_1,\dots,p_n) &= \underset{y_j\rightarrow p_j}{\text{DFT}}\left[ C_t^{\alpha_1,\dots,\alpha_n}(y_1,\dots,y_n) \right] \\ &\equiv   \sum_{y_1=1}^N\cdots\sum_{y_n=1}^N\text{e}^{-i2\pi y_1p_1}\cdots\text{e}^{-i2\pi y_np_n} C_t^{\alpha_1,\dots,\alpha_n}(y_1,\dots,y_n),
\end{align}
where $p_i  \in \left[ p_L,2p_L, \dots, Np_L \right]$, with $p_L = 1/L$ and $L$ the size of the evaluation region.
Further, we compared the results obtained with this procedure with an evaluation based on calculating the correlator in momentum space by Fourier transforming the single shot profiles and find no qualitative differences.

\subsection{Comparison with Gaussian fluctuations}

Correlation functions of order $>2$ will typically show non-vanishing values when inferred from finite statistics samples, even for the Gaussian case.
In order to check for significance we compare the correlations obtained from the experimental data to samples drawn from a Gaussian distribution with a covariance matrix given by the experimentally estimated two-point correlations. 
We analyse the profiles obtained from this routine in the same way as the experimental data.
The result of this procedure is shown as the upper limit of the grey shaded areas in Fig.~3 and Ext. Data Fig.~3.
Data points for which the 1~s.d. interval lies above this threshold are called significant.
For non-significant points we cannot make a statement whether they are zero in the infinite statistics limit or just too small to resolve their value with the amount of experimental realizations employed here.

\subsection{Equal-time formulation \label{equal_time_appendix}}
	We employ a formulation of quantum field theory based on equal-time quantities only \cite{Wetterich1997}.
	It is based on a generating functional $Z_t$, constructed from the density operator $\hat{\rho}_t$ as
	\begin{align}
		Z_t[J,J^*] = \text{Tr} \left[\hat{\rho}_t \,e^{ \int_x \left(J_x \hat{\psi}_x^\dagger + J_x^* \hat{\psi}_x \right)}\right] = \int\mathcal{D}\psi \mathcal{D}\psi^* \, W_t[\psi, \psi^*] \,e^{ \int_x \left(J_x \psi_x^* + J_x^* \psi_x \right)} \; .
	\end{align}
	$Z_t$ generates symmetrically ordered equal-time correlation functions of the two canonically conjugate fields $\hat{\psi}$, $\hat{\psi}^\dagger$ that fulfil $\left[\hat{\psi}_x,\hat{\psi}^\dagger_y\right]= \delta(x-y)$. Formally $Z_t$ is given by a functional integral with measure $\mathcal{D}\psi \mathcal{D}\psi^* = \prod_x \left[ \frac{1}{\pi} d \text{Re}\left(\psi_x\right) d \text{Im}\left(\psi_x\right)\right]$ over $W_t$, the field theoretic version of the	Wigner distribution which is extensively employed in quantum optics \cite{Hillrey1984}. 
	
	As described in Methods \textbf{B}, we consider the effective action $\Gamma_t$ corresponding to $Z_t$. 
	This construction implies the identity
	\begin{align}
		e^{-\Gamma_t[\Psi, \Psi^*]} = \int\mathcal{D}\psi \mathcal{D}\psi^* \, W_t[\psi, \psi^*] \, e^{  \int_x \left[J_x \left(\psi_x^* - \Psi^*_x\right)+J_x^* \left(\psi_x - \Psi_x\right)\right] }\,, 
	\end{align}
	where $J$,$J^*$ implicitly depend on $\Psi$, $\Psi^*$ due to the Legendre transform.
	Here, it is apparent that going from $W_t$ to $\Gamma_t$ represents a change of description from fluctuating  (microscopic) fields $\psi$ to averaged (macroscopic) fields $\Psi = \langle \hat{\psi} \rangle$.
	
	If $\hat{\rho}_t$ depends on no other degree of freedom than $\hat{\psi}$ and $\hat{\psi}^\dagger$, the knowledge of $\Gamma_t$ or $\hat{\rho}_t$ is fully equivalent. 
	Given the Hamiltonian $\hat{H}$ of the system, the von-Neumann equation is then equivalent to an evolution equation for $\Gamma_t$, i.e.
	\begin{align}\label{eq:evolution_equivalence}
		i \partial_t \hat{\rho}_t = \left[\hat{H},\hat{\rho}_t\right]	&& \Leftrightarrow && i \partial_t \Gamma_t = \mathcal{L} \Gamma_t \; .
	\end{align}
	Here $\mathcal{L}$ \citep{Wetterich1997} is functional-differential operator that depends on the form of $\hat{H}$.
	Taking functional derivatives of the evolution equations for $\Gamma_t$ in Eq. \eqref{eq:evolution_equivalence} results in a hierarchy of coupled evolution equations in terms of equal-time 1PI correlators. 
	This hierarchy is a non-equilibrium, equal-time version of the Schwinger-Dyson equations and constitutes a field theoretic analog of the quantum BBGKY hierarchy. 
	Considering a non-relativistic Bose gas of particles with mass $m$ and interaction strength $g$, described by
	\begin{align}
		\hat{H} = \int_x \left[\frac{1}{2m} \left(\nabla_x \hat{\psi}_x^\dagger\right) \left(\nabla_x \hat{\psi}_x\right) + \frac{{g}}{2} \hat{\psi}^\dagger_x\hat{\psi}^\dagger_x\hat{\psi}_x\hat{\psi}_x\right] \;,
	\end{align}
	we obtain the evolution equation for the two-point function given in Eq. (3). 
	It is formally exact and assumes only spatially translation invariant and $U(1)$ symmetric correlators.

\vspace*{0.5cm}
\newpage
\begin{table}[h!]
  \begin{center}
    \begin{tabular}{|c|c|} 
    \hline
      \textbf{Evolution Time (s)} & \textbf{Number of realizations}\\
      9 & 294 \\
      12 & 511 \\
      15 & 316 \\
      18 & 559 \\
      \hline
    \end{tabular}
    
    \caption{Number of experimental realizations for the evolution times shown in the main text.}
  \end{center}
\end{table}

\vspace*{0.5cm}
\newpage
\begin{center}
  \textbf{Extended Data Figures}
\end{center}
\setcounter{figure}{0}

\vspace*{0.5cm}

\begin{figure}[!ht]
	\centering
	\includegraphics[width = 1\textwidth]{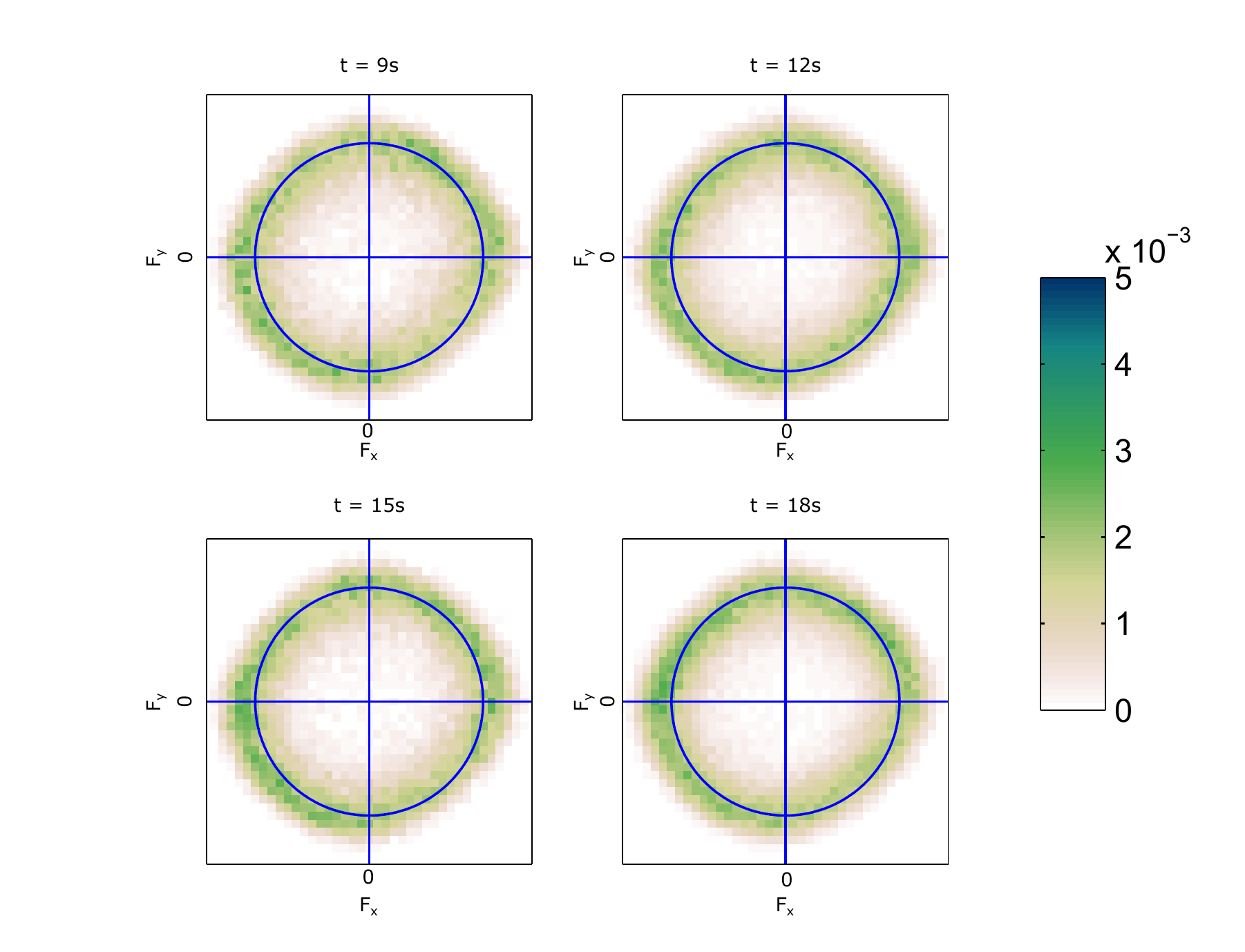}
	\caption{\textbf{Observation of the complex valued field for all times.} We show the distribution of $F_\perp$ (as in Fig.~1c) for all evolution times shown in Fig.~4. The blue circle indicates $|F_\perp| = 0.7$.}
	\label{MeanAbsorption150}
\end{figure}

\begin{figure}[!ht]
	\centering
	\includegraphics[width = 1\textwidth]{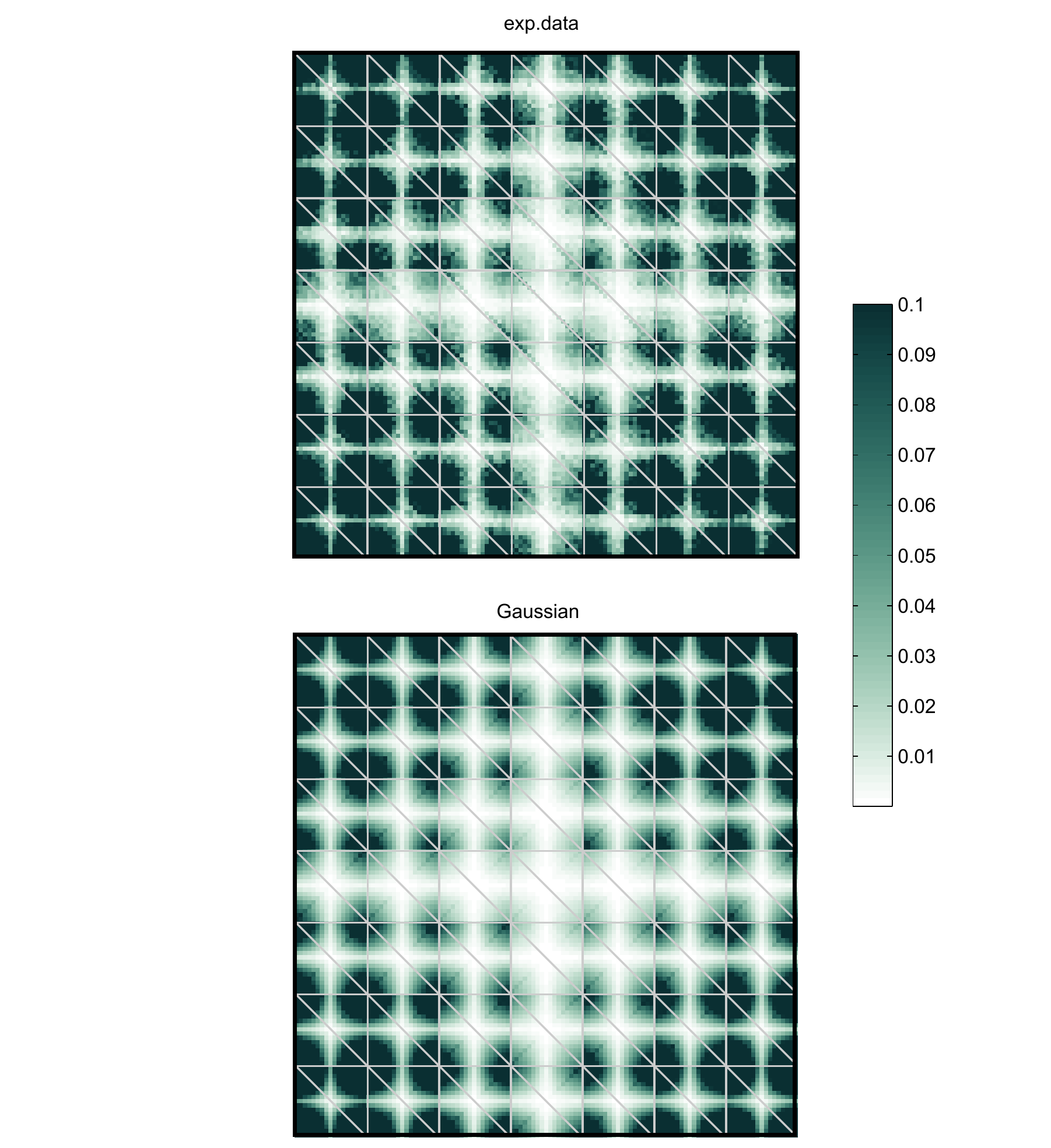}
	\caption{\textbf{1PI four vertex.} $\Gamma_t^{(4)}(p_1,p_2,p_3,p_4)$ inferred from the data for the same momenta as shown in Fig.~2c. For comparison we show the results for the Gaussian model.}
	\label{MeanAbsorption150}
\end{figure}

\begin{figure}[!ht]
	\centering
	\includegraphics[width = 1\textwidth]{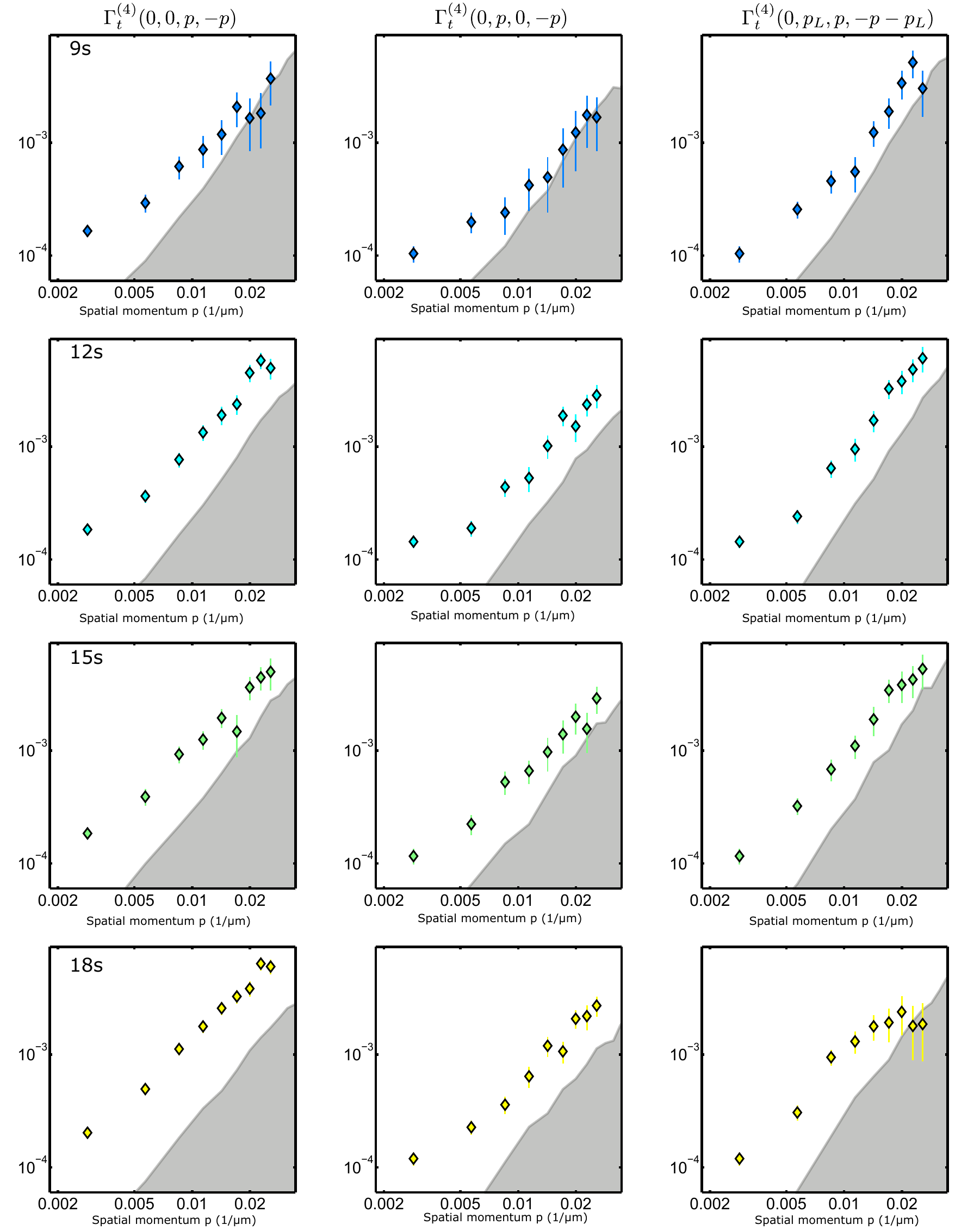}
	\caption{\textbf{Momentum conserving diagonals.} Momentum conserving diagonal of the four-point 1PI correlators shown in Fig.~3 for all evolution times between 9\,s and 18\,s. Grey shaded area is the finite statistical bias from a Gaussian model.}
	\label{MeanAbsorption150}
\end{figure}

\begin{figure}[!ht]
	\centering
	\includegraphics[width = 1\textwidth]{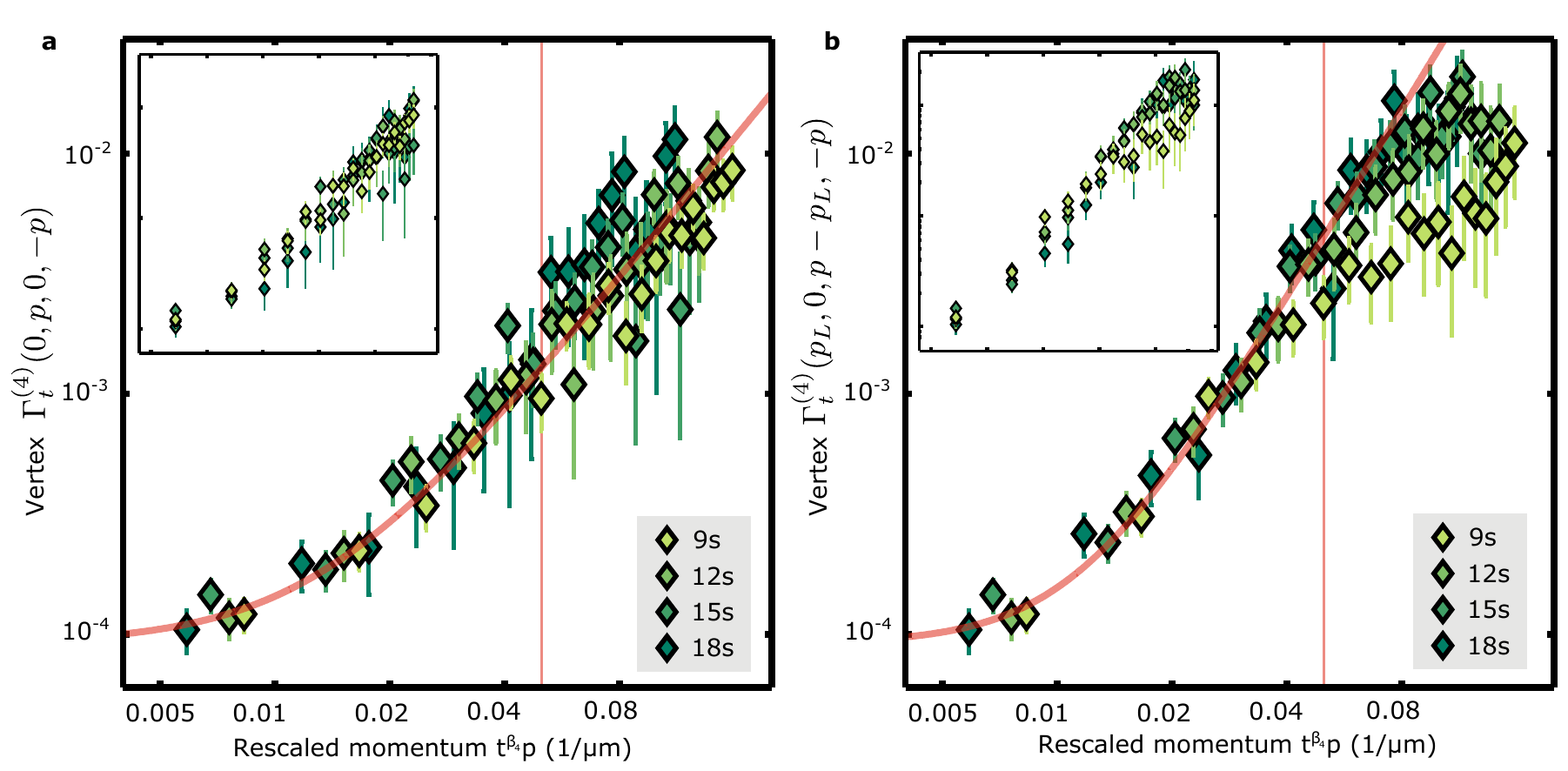}
	\caption{\textbf{Rescaling of different momentum conserving diagonals.} Momentum conserving diagonals of the 1PI four-point correlator shown in Fig.~3 for all evolution times between 9\,s and 18\,s rescaled as in Fig.~4b.}
	\label{MeanAbsorption150}
\end{figure}

\clearpage

\bibliography{../main}

\newpage
\noindent\textbf{Acknowledgements}\\
We thank Sebastian Erne, Thomas Gasenzer, Philipp Hauke, Christian-Marcel Schmied, J\"org Schmiedmayer, Thomas Schweigler and Malo Tarpin for discussions and Rodrigo Rosa-Medina for experimental assistance.\\
This work was supported by the DFG Collaborative Research Center SFB1225 (ISOQUANT) and the ERC Advanced Grant Horizon 2020 EntangleGen (Project-ID 694561), by Deutsche Forschungsgemeinschaft (DFG) under Germany's Excellence Strategy EXC-2181/1 - 390900948 (the Heidelberg STRUCTURES Excellence Cluster) and the Heidelberg Center for Quantum Dynamics. P.K. acknowledges support from the Studienstiftung des deutschen Volkes.\\
\noindent\textbf{Author contributions}, 
The experimental and theoretical concept was developed in discussion among all authors.  M.P., P.K., S.L., and A.B. controlled the experimental apparatus. M.P. and T.V.Z. analysed the data. M.P., T.V.Z., P.K., H.S., J.B., and M.K.O. discussed the measurement results. T.V.Z. and J.B. elaborated the equal time formalism. All authors contributed to the discussion of the results and the writing of the manuscript.

\noindent\textbf{Competing financial interests}\\
The authors declare no competing financial interests.

\end{document}